\documentstyle[12pt]{article}

\begin{document}
\begin{flushright}
FSU--HEP--931026\\
hep-ph/9310351\\
October 1993
\end{flushright}
\begin{center}
{\large\bf Measuring the Gluon Helicity Difference Distribution
Function of the
Proton using Photoproduction Processes
\footnote{to appear in the
proceedings of the International Europhysics Conference on
High Energy Physics, Marseille, July 22--28 1993}
}

\vspace{.7in}

{S.~Keller}

\vspace{.55in}

{ \it Department of Physics, B-159, Florida State University\\
 Tallahassee, Florida 32306, U.S.A\\}
\end{center}
\begin{abstract}\normalsize\baselineskip24pt

Little information is known about the polarization of gluons inside  a
longitudinally polarized proton. I report on the sensitivity of
photoproduction experiments to it.  Both jet and heavy quark production
are considered.

\end{abstract}

{\bf \noindent Introduction}
\vskip 2.5mm
Since the so-called EMC spin crisis has emerged\cite{EMC88}, much
experimental and theoretical work has been done\cite{Bas92}.
One remaining question is the size
of the gluon helicity difference distribution function ($\Delta g$).
In this contribution, the sensitivity to $\Delta g$
is studied in photoproduction experiments where both the photon
and the proton are longitudinally polarized.  The photoproduction of jets and
heavy quarks is considered.
As is well know,  photoproduction processes receive contributions
from two classes  of subprocesses.  In the first class, the
photon
interacts directly with the constituents of the proton (the ``direct''
contribution).  In the second class, the photon interacts through its
distribution functions (the ``resolved'' contribution).
For the unpolarised distribution functions of the proton (photon), the set
DO1.1~\cite{Owe91} (D0~\cite{DO}) is used.  For the helicity difference
distribution
functions of the proton the three sets (set 1, 2, and 3) developed in
Ref.~\ref{Kel93} are used.
At the initial $Q_0^2=4\  {\rm GeV}^2$, the three sets have identical
quark helicity difference
distribution function, but different $\Delta g$, see Fig.~\ref{fig:protdist}.
Clearly, the three sets can be used to study the sensitivity of an observable
to $\Delta g$.
The parametrization of Ref.~\ref{Has81} is used for the helicity difference
distribution function of the photon.

\vskip 5.4mm
{\bf\noindent Two-jet production}
\vskip 2.0mm

One observable sensitive to $\Delta g$ is the longitudinal
asymmetry, defined as:
\begin{eqnarray}
\nonumber \\
A_{ll}={\sigma^{++}-\sigma^{+-}\over \sigma^{++}+\sigma^{+-}}.\\
\nonumber
\label{eq:All}
\end{eqnarray}
where $\sigma^{++}$ ($\sigma^{+-}$ ) is the cross section for same (opposite)
sign helicity of the photon and proton.
The asymmetry for the diret contribution is presented in
Table~\ref{tab:jasymdir}, at
$E_\gamma=200\ {\rm GeV}$ and
$p_{\rm T}(jet) \geq 3 \ {\rm GeV}$, for set 1 and 3 (smallest and largest
$\Delta g$).  $E_\gamma=200\ {\rm GeV}$ corresponds to the average value for
$E_\gamma$ of  present unpolarized  experiments and
$p_{\rm T}(jet)=3\ {\rm GeV}$ is the lowest value
at which jets have been observed in fixed target  experiments\cite{Cor92}.
Also shown in Table~\ref{tab:jasymdir} are the asymmetries for the quark and
gluon contributions corresponding to subprocesses involving a quark or a gluon
inside the proton, respectively.
\begin{figure}
\vspace{2.5in}
\caption[fig1]
{Gluon helicity sum (solid) and helicity difference (dashes) distribution
functions of the proton at $Q_0^2=4\ {\rm GeV}^2$  for set 1 (lower),
2 (middle) and 3 (upper).
}
\label{fig:protdist}
\end{figure}
\begin{table}
\centering
\begin{tabular}{|c|c|c|c|} \hline
&  quark & gluon& direct\\ \hline
set 1 &25. &-8.3    &7.8	      \\	\hline
set 3 &25. &-93.    &-36.	      \\	\hline
\end{tabular}
\caption
[Asymmetries for dijet production.]
{Asymmetries (\%) of the direct contribution for dijet production for set
1 and 3, $E_\gamma=200\ $GeV, and $p_{\rm T}(jet) \geq 3\ $GeV.
}
\label{tab:jasymdir}
\end{table}
The quark contribution gives a positive asymmetry and there is no difference
between the two sets.  The gluon contribution is negative and, as expected, the
difference between the two sets is large, of the order of 85\%.  The cross
section of the quark and gluon contribution are about equal at this energy,
such that the difference between the two sets for the direct contribution is
about half of the difference for the gluon contribution, $\sim 40\%$.
The total asymmetry is presented in Table~\ref{tab:jasymtot},
along with the asymmetry of the direct and resolved contribution.  The
difference between the two sets in the total asymmetry is only about 15\%.
The problem stems from the fact that the gluon contribution is negative in the
direct case and positive in the resolved case, such that the two contributions
partially cancel each other.  An obvious way to improve upon
this is to separate the direct and resolved contributions, and then  use the
direct contribution to measure $\Delta g$, as it is the most sensitive
contribution.  The
same techniques developed for the unpolarized case can be implemented to
separate the direct and resolved contributions~\cite{Dre89}.
\begin{table}
\centering
\begin{tabular}{|c|c|c|c|} \hline
&  diret& res&total\\ \hline
set 1 &7.8 &2.7    &5.3	      \\	\hline
set 3 &-36. &17.    &-10.	      \\	\hline
\end{tabular}
\caption
[Asymmetries for dijet production.]
{Asymmetries (\%) for dijet production for set 1 and 3 , $E_\gamma=200\ $GeV,
and $p_{\rm T}(jet)\geq 3\ $GeV.
}
\label{tab:jasymtot}
\end{table}

\begin{figure}
\raggedright
\vspace{2.5in}
\caption[fig2]
{Asymmetry distribution of the direct contribution for set 1 (dashes),
set 2 (dots), and set 3 (solid) as a function of $log_{10}(x_p)$.
}
\label{fig:jxp200}
\end{figure}

More detailed information can be obtained by looking at the longitudinal
asymmetry of the differential cross sections.  In Fig.~\ref{fig:jxp200} the
$x_p$-distribution of the direct contribution is presented as a representative
example.

\vskip 4.0mm
{\bf \noindent Heavy Quark production}
\vskip 2.0mm

As is well known, the resolved contribution for the photoproduction
of heavy quarks for the energy range considered here  is of the order of a few
percent, and can be neglected.
It turns out that the asymmetry is positive in some regions of phase space and
negative in others~\cite{fig8}.  Therefore, care must be taken when
trying to evaluate
the sensitivity of heavy quark production to $\Delta g$; it is
bigger than suggested by the integrated asymmetry.
\vskip 4.0mm
{\bf \noindent Conclusions}
\vskip 2.0mm

Considering the total asymmetry, one can show that two
jet and heavy quark production have similar sensitivities to $\Delta g$.
The best way to
measure the gluon helicity difference  distribution function is by using two
jet production at low $p_{\rm T}(jet)$, with separation of direct and resolved
contribution, and then to use the direct contribution which has the biggest
sensitivity.
\newpage
{\bf \noindent Acknowledgements}
\vskip 2.0mm
The author thanks J.~F.~Owens for collaboration on this
topic~\cite{Kel93}.  This research was supported in part by the Texas
National Research
Laboratory Commission and by the U.S. Department of Energy under contract
number DE--FG05--87ER40319.


\begin{thebibliography}{99}

\bibitem{EMC88}\label{EMC88}
J.Ashman {\it et al.} , EMC collaboration, {\it Phys.\ Lett.\ }{\bf 206B}
(1988) 364;  {\it Nucl.\ Phys.\ }{\bf B328} (1989) 1.


\bibitem{Bas92}\label{Bas92}
For an overview see S.~D.~Bass and A.~W.~Thomas, CAVENDISH--HEP--92--5,
ADP--92--1833--T115, SMC--92--25, Aug. 1992,
and reference therein.

\bibitem{Kel93}\label{Kel93}
S.~Keller and J.~F.~Owens, preprint FSU-HEP-930609, to be published in {\it
Phys. Rev.} {\bf D}.

\bibitem{Has81}\label{Has81}
J.~A.~Hassan and D.~J.~Pilling, {\it Nucl.\ Phys.\ }{\bf B187} (1981) 563.

\bibitem{Owe91}\label{Owe91}
J.~F.~Owens, {\it Phys.\ Lett.\ }{\bf 266B} (1991) 126.


\bibitem{DO}\label{DO}
D.~W.~Duke and J.~F.~Owens, {\it Phys.\ Rev.\ }{\bf D30}
(1984) 49.



\bibitem{Cor92}\label{Cor92}
M.~Corcoran, private communication.


\bibitem{Dre89}\label{Dre89}
M.~Drees, and R.~M.~Godbole,
{\it Phys.\ Rev.\ }{\bf D39} (1989) 169;  R.~S.~Fletcher, F.~Halzen,
S.~Keller, and W.~H.~Smith, {\it Phys.\ Lett.\ }{\bf 266B} (1991) 183.

\bibitem{fig8}
see Fig. 8 in Ref.~\ref{Kel93}.

\end{thebibliography}
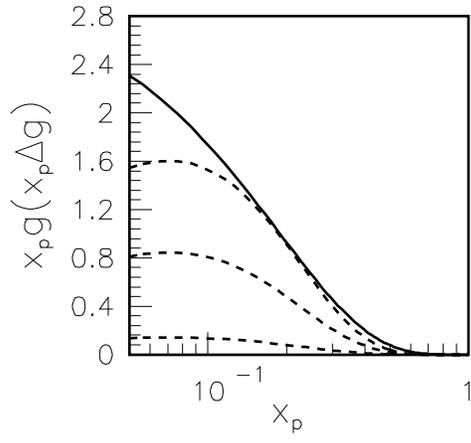
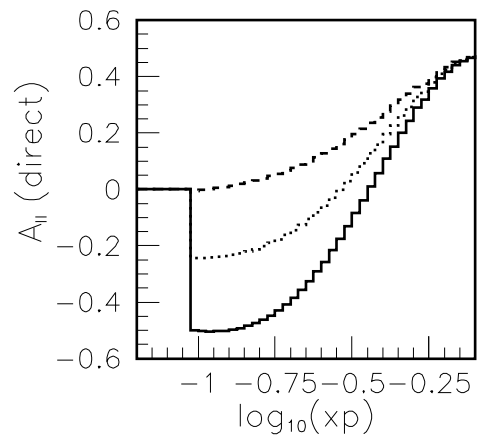
\end{document}